# THE GREAT INEQUALITY IN A HAMILTONIAN PLANETARY THEORY

F. Varadi[1], M. Ghil[1,2] and W. M. Kaula[1,3]

[1]Institute of Geophysics and Planetary Physics
[2]Department of Atmospheric Sciences
[3]Department of Earth and Space Sciences
 University of California, Los Angeles
 Los Angeles, CA 90024

**Abstract.** The Jupiter–Saturn 2:5 near–commensurability is analyzed in a fully analytic Hamiltonian planetary theory. Computations for the Sun–Jupiter–Saturn system, extending to the third order of the masses and to the 8th degree in the eccentricities and inclinations, reveal an unexpectedly sensitive dependence of the solution on initial data and its likely nonconvergence. The source of the sensitivity and apparent lack of convergence is this near–commensurability, the so–called great inequality. This indicates that simple averaging, still common in current semi–analytic planetary theories, may not be an adequate technique to obtain information on the long-term dynamics of the Solar System. Preliminary results suggest that these difficulties can be overcome by using resonant normal forms.

## INTRODUCTION

The long–term stability of the Solar System is one of the oldest unsolved problems in classical mechanics (Duncan and Quinn, 1993). Recent studies on the motion of the major planets use a variety of techniques. They range from the purely analytical (Duriez, 1979) through the semi–analytical (Laskar 1985, 1988; Wisdom and Holman, 1991) to the purely numerical (Carpino *et al.*, 1987; Applegate *et al.*, 1986; Quinn *et al.*, 1991). Our theory (Varadi and Ghil, 1993a,b) follows the ideas of Message (1982, 1988); it is analytic and fully Hamiltonian.

The dynamics of the Sun–Jupiter–Saturn system was recognized as problematic from the beginnings of perturbation theory. The problems are due to the so–called Great Inequality (GI) which is the Jupiter–Saturn 2:5 mean–motion near–commensurability. Brouwer and Van Woerkom (1950) (see also Knezevic, 1986), being aware of this, included some extra terms in their expansions, trying to account for the effects of the GI. We wanted to test our theory on this undoubtedly difficult case, and were interested in any signs of nonconvergence. Alternatively, in the case of apparent convergence, one would like to know the appropriate truncation of the expansions.

In some recent numerical integrations, evidence of chaos – defined as the presence of a positive Lyapunov exponent – has been found in the motion of the major planets (Laskar, 1990; Laskar *et al.*, 1992; Sussman and Wisdom, 1992). There is no consensus yet regarding the exact nature of this chaos or the unpredictability of planetary motion it might entail. It is natural to ask whether the GI is at least partly responsible for the apparently chaotic behavior of the Jovian planets.

Our computations extended to the 3rd order in the masses and to the 8th degree of eccentricities and inclinations. To our surprise we obtained vastly different values of the orbital parameters for small changes in the initial data. The expansions themselves are apparently nonconvergent. We repeated the computations without the GI terms. As a result, the sensitive dependence disappears and the expansions appear to converge. This is a positive result: perturbation theory does work for small perturbations. Averaging, however, as it is normally carried out using Lie–series or other methods for the necessary transformations in both Hamiltonian and non–Hamiltonian theories, is not adequate to deal with the GI.

One might infer from this evidence that the GI generates, through some yet unknown mechanism, chaotic motion. We do not think that this is the case since perturbation theory has its limitations. In order to illustrate these limitations, we present a simple example when a certain type of perturbation theory leads to the wrong conclusion.

We also have evidence that the GI can be dealt with existing techniques based on resonant normal forms. Preliminary computations, based on earlier results (Varadi, 1989), indicate that the nonconvergent behavior is not present when the appropriate normal form is used.

## AN EXAMPLE OF AN INADEQUATE PERTURBATION THEORY

In order to put our GI results into the proper perspective, we present an example where classical time–dependent perturbation theory fails. This is intended to demonstrate that the failure of a particular perturbation theory does not necessarily mean that the problem cannot be solved by means of a better theory.

The issue of secular terms in the semi–major axes in non–Hamiltonian planetary theories occupied many researchers during the last century. We think that our deceptively simple example is a good demonstration of the basics of the phenomenon, without getting lost in the technical details that arise in actual planetary computations.

We applied the method of successive approximations (i.e., Picard iteration) used in classical perturbation theory (Brouwer and Clemence, 1961) to the case of a pendulum subject to a small gravitational force. The Hamiltonian is $p^2/2 + \epsilon \sin q$. The zeroth–order solution is $q = t$, $p = 1$. The third–order solution is

$$q = t + \epsilon(-1 + \cos t) + O(\epsilon^4) , \tag{1a}$$

$$p = 1 - \epsilon \sin t + \epsilon^2 \left(-\frac{1}{2} + \cos t - \frac{\cos t^2}{2}\right) +$$

$$\epsilon^3 \left(-\frac{t}{2} + \frac{7 \sin t}{8} - \frac{\sin 2t}{4} + \frac{\sin 3t}{24}\right) + O(\epsilon^4) . \tag{1b}$$

There is a secular term of type $\epsilon^3 t$ in the solution. This is clearly due to the perturbation method used, since $p$ remains bounded in the actual system. The secular term appears in the third–order solution first, i.e., at the same order as in the classical perturbation theory of planetary motions.

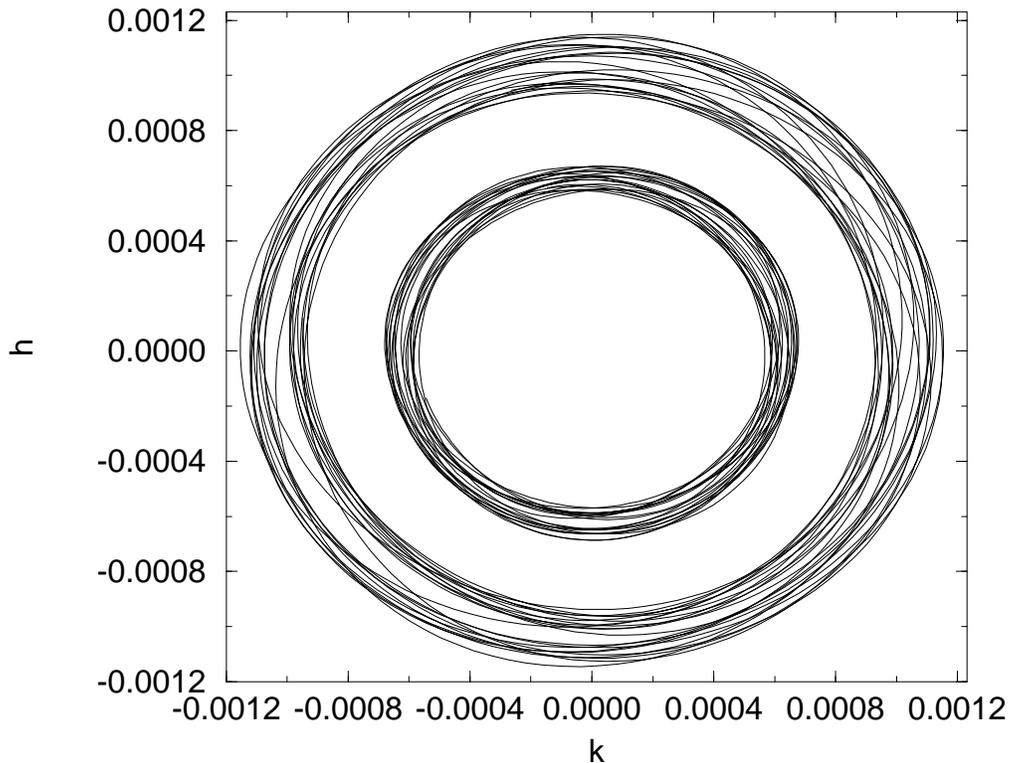

**Figure 1.** Jupiter's reduced $(h, k)$ coordinates for runs A (inner curve) and Q (outer curve); see text for abbreviations.

## THE EFFECTS OF THE GREAT INEQUALITY

The Poisson–series processor described in Varadi and Ghil (1993b) has been replaced by a more efficient one. In a typical run for the Sun–Jupiter–Saturn computation at hand, which takes about one day on a Sparc workstation, we are able to compute the Poisson bracket of two expansions having half a million terms each.

For the sake of brevity we use the letter O to refer to order in the masses, and the letter D to refer to degree in eccentricities and inclinations. The secular Hamiltonian includes all terms up to O1D8, O2D6 and O3D4. We call the result of the Birkhoff normalization of this secular system the full solution.

We used two sets of initial data. The first one, labeled A, is the initial data of the Applegate *et al.* (1986) integration, the second one, Q, is that of Quinn *et al.* (1991). These two are supposed to correspond to the same Solar System at slightly different epochs, i.e., most of the difference comes from different phases in the short–periodic perturbations. The transformation of initial data – necessitated by the use of Lie series (Henrard, 1970; Varadi, 1993) – was carried out only in the Birkhoff normalization of the secular system.

**Table 1.** Comparison accross runs.

|  | A | Q | AZ | QZ |
|---|---|---|---|---|
| Jupiter's semi–major axis | 100.000 | 99.968 | 100.000 | 99.968 |
| Saturn's semi–major axis | 100.000 | 100.078 | 100.000 | 100.078 |
| Period of GI | 100.000 | 138.857 | 100.000 | 138.857 |
| Frequency of $f_5$ at O1D2 | 100.000 | 99.775 | 100.000 | 99.775 |
| Frequency of $f_5$ at O2D6 | 100.000 | 103.357 | 95.407 | 95.141 |
| Frequency of $f_6$ at O2D6 | 100.000 | 112.950 | 81.520 | 81.195 |
| Frequency of $g_6$ at O2D6 | 100.000 | 99.645 | 99.900 | 99.542 |
| Amplitude of $f_5$ in J(h,k) | 100.000 | 100.533 | 97.968 | 97.225 |
| Amplitude of $-f_5 + 2f_6$ in J$(h, k)$ | 100.000 | 165.251 | 9.125 | 9.474 |
| Amplitude of $-f_5 + 2f_6$ in S$(h, k)$ | 100.000 | 168.886 | 6.637 | 6.850 |
| Amplitude of $-2f_5 + 3f_6$ in J$(h, k)$ | 100.000 | 211.582 | 0.381 | 0.411 |

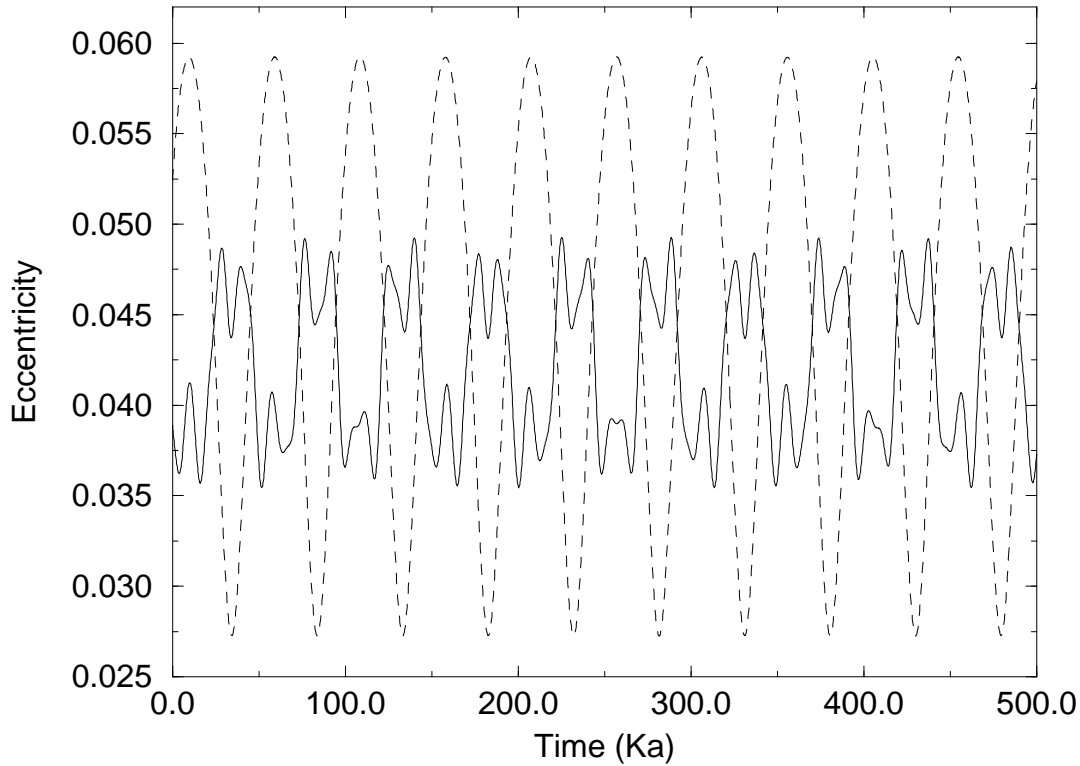

**Figure 2.** Jupiter's total (dashed line) and reduced (solid line) eccentricity for run Q.

One advantage an analytic theory has over numerical integration is that one can eliminate certain terms from the expansions. We carried out computations where all terms related to the GI were eliminated. We refer to these calculations as runs AZ and QZ, respectively.

Solutions for the Jovian planets were computed at various levels of truncation. For initial data A we obtained the same results as in Varadi and Ghil (1993a,b), at the same level of truncation. In the light of the present results, it appears that we were lucky; the truncation happened to be such that the results were relatively close to the results of numerical integrations.

Our results are summarized in two tables and three figures. We use the classical notation of Brouwer and Clemence (1961), i.e., $f$ refers to the frequency of the longitude of perihelion, while $g$ refers to that of the ascending node.

Table 1 compares the solutions we obtained. The numbers are normalized with respect to the appropriate entries in the first column. The amplitudes are for the full solution. A 0.19% change in initial, instantaneous semi–major axes can cause about 40% change in the period of the GI. For runs A and Q this leads to a 13% change in $f_6$. The amplitude of the $-f_5 + 2f_6$ term, which is the largest combination tone of the fundamental frequencies, changes by over 60% between A and Q. The smaller combination terms show even bigger differences, e.g., $-2f_5 + 3f_6$ differs by 111%. The cause of the large differences is obviously the GI, since the differences between AZ and QZ are much smaller and of the expected order for regular perturbations.

**Table 2.** Values of the secular Hamiltonian for runs Q and QZ.

|  | Degree 2 | | Degree 4 | | Degree 6 | | Degree 8 | |
|---|---|---|---|---|---|---|---|---|
|  | Q | QZ | Q | QZ | Q | QZ | Q | QZ |
| Order 1 | 100.000 | 100.000 | 0.5796 | 0.5796 | 0.3156e-2 | 0.3156e-2 | 1.0845e-4 | 1.0845e-4 |
| Order 2 | 108.169 | 108.169 | 8.3584 | 0.8050 | 4.7429 | 0.3594e-2 | | |
| Order 3 | 108.172 | 108.306 | 4.2899 | 0.7936 | | | | |

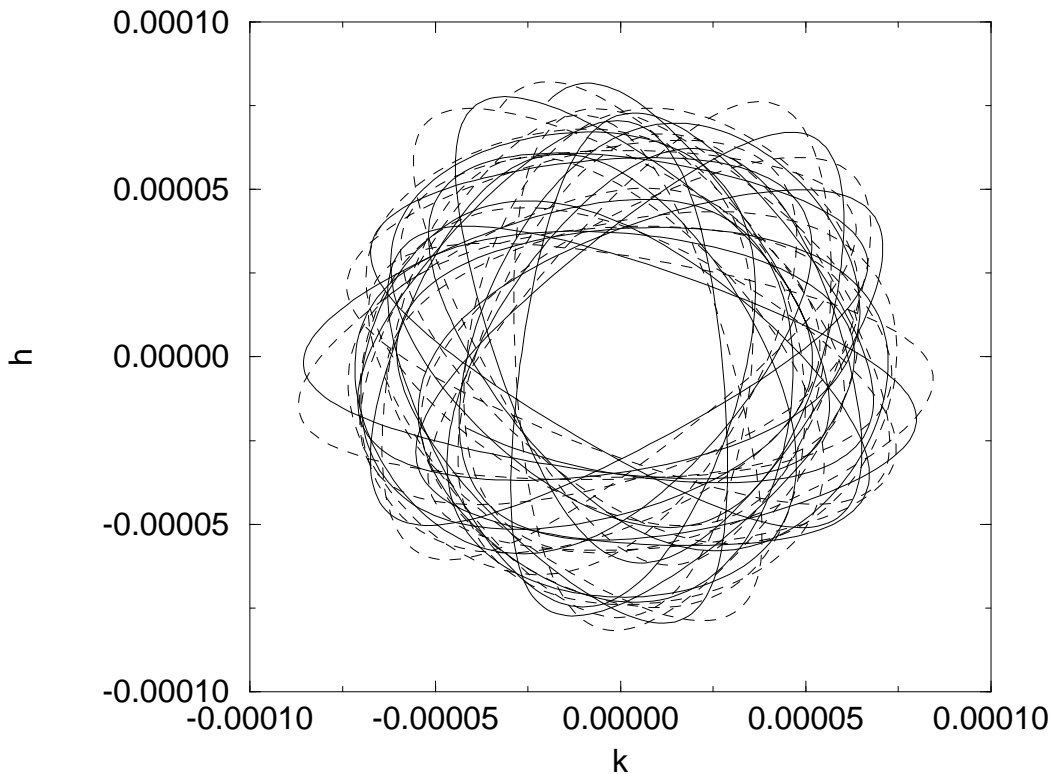

**Figure 3.** Jupiter's reduced $(h, k)$ for runs AZ (solid line) and QZ (dashed line).

Table 2 lists the values of the secular Hamiltonian for Q and QZ. For each order we list the contribution of the various degrees; the numbers are normalized with respect to the value for O1D2, the classical Lagrange–Laplace secular system. The case of QZ shows what one would expect from a convergent expansion: rapidly decreasing values with respect to degree and small changes with respect to order. The case of Q shows none of these. The value of the secular Hamiltonian changes by 4% between O2 and O3 for the D4 terms and by another 4% between O2D4 and O2D6. The Lie–series generator for the Birkhoff normalization will obviously behave even worse. In fact, its value for the O2D6 terms is larger than for the O2D4 terms (not shown).

To visualize the nature of these differences, we plotted the solution for the variables $(h = e \sin \varpi, k = e \cos \varpi)$ over 500 thousand years in Figures 1–3, $e$ being the eccentricity and $\varpi$ the longitude of perihelion. When one plots the full solutions (not shown) the difference in the fundamental frequencies, i.e., $f_5$ and $f_6$, is clearly visible. We define as a reduced solution one omitting the $f_5$ and $f_6$ terms. The reduced solution curves are dominated by the $-f_5 + 2f_6$ term, so they exhibit the relatively narrow rings in Figure 1. The difference in the amplitudes of the two reduced solutions clearly generates two distinct rings. The picture is reminiscent of period–doubling bifurcation but we have not verified this to be the case.

Figure 2 shows how the fundamental terms and the reduced solution are related. We plotted the total eccentricity and the reduced solution's eccentricity for 500 thousand years; the latter is magnified by about a factor of 50 for better visualization. The reduced solution has a high and a low eccentricity state with regular and fast transitions between them. Its phase is opposite to that of the fundamental terms.

The behavior of the AZ and QZ solutions in Figure 3 shows only differences of phase and not of amplitudes. The scale of this plot is about one–tenth of that in Figure 1. Moreover, the solutions are not dominated by the $-f_5 + 2f_6$ term. Because of lack of space we are not able to show The eccentricity plot for AZ and QZ (not shown) does not exhibit the same bimodality as Figure 2, although the reduced solution still tends to counteract the fundamental terms.

Having shown that simple averaging does not work for the Sun–Jupiter–Saturn system, our next task is to carry out the details of a resonant normal form (Varadi, 1989) for the GI. Our preliminary results indicate that an apparently convergent theory can be obtained for all four Jovian planets. This brings us back to the quandary mentioned in the introduction: chaos in numerical integrations, on the one hand, and quasiperiodicity in an analytic theory, on the other, make for a very puzzling scenario, indeed.

**Acknowledgments.** The financial support of NSF Grant ATM90–13217 (M. Ghil and F. Varadi), NASA Grant NAGW–2269 (W. M. Kaula and F. Varadi) and a Guggenheim Fellowship (M. Ghil) is gratefully acknowledged.

# REFERENCES

Applegate, J. H., Douglas, M. R., Gursel, Y., Sussman, G. J. and Wisdom, J., 1986, The outer Solar System for 200 million years, *Astron. J.* 92:176

Brouwer, D. and Clemence, G. M., 1961, "Methods of Celestial Mechanics", Academic Press, Orlando

Brouwer, D. and Van Woerkom, A. J. J., 1950, The secular variations of the orbital elements of the principal planets, *Astron. P. Amer. Eph.* 13(2):81

Carpino, M., Milani, A., and Nobili, A. M., 1987, Long–term numerical integrations and synthetic theories for the motion of the outer planets, *Astron. Astrophys.* 181:182

Duncan, M. J. and Quinn, T., 1993, The long–term dynamical evolution of the Solar System, *Annu. Rev. Astron. Astrophys.* 31:265

Duriez, L., 1979, Approche d'une théorie général planétaire en variables elliptiques héliocentriques, Thèse de doctorat, Lille

Henrard, J., 1970, On a perturbation theory using Lie transforms, *Celest. Mech.* 3:107

Knezevic, Z., 1986, Secular variations of the major planets' orbital elements, *Celest. Mech.* 38:123

Laskar, J., 1985, Accurate methods in general planetary theory, *Astron. Astrophys.* 144:133

Laskar, J., 1988, Secular evolution of the Solar System over 10 million years, *Astron. Astrophys.* 198:341

Laskar, J., 1990, The chaotic motion of the Solar System: a numerical estimate of the size of the chaotic zones, *Icarus* 88:266

Laskar, J., Quinn, T. and Tremaine, S., 1992, Confirmation of resonant structure in the Solar System, *Icarus* 95:148

Message, P. J., 1982, Asymptotic series for planetary motion in periodic terms in three dimensions, *Celest. Mech.* 26:25

Message, P. J., 1988, Planetary perturbation theory from Lie series, including resonance and critical arguments', *in* "Long–Term Dynamical Behaviour of Natural and Artificial N–body Systems", A. E. Roy, ed., Kluwer Academic Publ., Dordrecht

Quinn, T. Q., Tremaine, S. and Duncan, M., 1991, A three million year integration of the Earth's orbit, *Astron. J.*, 101:2287

Sussman, G. J. and Wisdom, J., 1992, Chaotic evolution of the Solar System, *Science*, 257:56

Varadi, F., 1989, Hamiltonian perturbation theory applied to planetary motions, Ph.D. Thesis, University of California, Los Angeles

Varadi, F., 1993, Branching solutions and Lie series, *Celest. Mech. Dyn. Astron.*, 57:517

Varadi, F. and Ghil, M., 1993a, Hamiltonian planetary theory. Part I: Theoretical preliminaries, submitted to *Celest. Mech. Dyn. Astron.*

Varadi, F. and Ghil, M., 1993b, Hamiltonian planetary theory. Part II: The secular system and numerical results, submitted to *Celest. Mech. Dyn. Astron.*

Wisdom, J. and Holman, M., 1991, Symplectic maps for the *n*–body problem, *Astron. J.*, 102:1528